\documentclass[fleqn,11pt]{article}
\usepackage{amsmath}
\usepackage{amsfonts}
\usepackage{rotate,epsfig}
 \usepackage{lscape}
\DeclareMathSizes{1}{5}{5}{5}

    {\par \noindent {\bf Proof:}}%
    {\par \indent}

\usepackage{epsfig}
\usepackage{amssymb}
\usepackage{fancyhdr}
\usepackage{amsmath}

\newenvironment{algorithm}%
    {\par \noindent {\bf Algorithm I:}}%
    {\par \indent}

\begin{document}

\title{Optimal design for step-stress accelerated test
with random discrete stress elevating times based on \\gamma degradation process}

\author{Morteza Amini$^\dag$,\footnote{Corresponding author, E-mail addresses: morteza.amini@ut.ac.ir, shemehsavar@khayam.ut.ac.ir, panzhq.yy@hotmail.com}~~  Soudabeh Shemehsavar$^\dag$ ~ and ~ Zhengqiang Pan$^\ddag$\\
{\small $\dag$ Department of Statistics, School of
Mathematics, Statistics and Computer Sciences,}\\
{\small College of Science, University of Tehran,
P.O. Box 14155-6455, Tehran,Iran}\\
{\small $\ddag$ Department of Systems Engineering, 
College of Information Systems and Management, }\\
{\small National University of Defense Technology, 
Changsha, People's Republic of China}}
\maketitle

\begin{abstract}
Recently, a step-stress accelerated degradation test (SSADT) plan,
in which the stress level is elevated when the degradation value of a product
crosses a pre-specified value, was proposed. The times of stress level elevating are
random and vary from product to product. In this paper we extend this model to a more economic plan.
The proposed extended model has two economical advantages compared with the previous one.
The first is that the times of stress level elevating in the new model are
identical for all products, which enable us to use only one chamber (oven) for testing all test units. The second is
that, the new method does not require continuous inspection and to elevate the stress level, it is not necessary for the experimenter
to inspect the value of the degradation continually.
The new method decrease the cost of measurement and also there is no need to
use electronic sensors to detect the first passage time of the degradation to the threshold value in the new method. We assume that
the degradation path follows a gamma process. The stress level is elevated as soon as the measurement
of the degradation of one of the test units, at one of the specified times, exceeds the threshold value. Under the
constraint that the total experimental cost does not exceed a
pre-specified budget, the optimal settings including the optimal
threshold value, sample size, measurement frequency and termination
time are obtained by minimizing the asymptotic variance of an
estimated quantile of the lifetime distribution of the product.
A case study is presented to illustrate the proposed method.
\end{abstract}

\noindent{\bf Keywords:} Discrete times, Fisher information matrix, Gamma process, Inverse Gaussian distribution.

\section{Introduction}

Due to the rapid improvement of quality of
products of today, it is difficult to assess the failure information of a
product using traditional life testing procedures, which record only
failure times. An alternative approach is to collect the degradation
data at higher levels of stress, thus yielding more information about
the lifetime distribution in a reasonable amount of time compared to
traditional life tests. Such a life testing plan is called an
accelerated degradation test (ADT). Mathematical models for
analyzing ADT data have been proposed by Meeker \& Escobar \cite{me}.
Meeker \textit{et al.} \cite{mel} described accelerated degradation
models that relate to physical failure mechanisms. Yu \& Tseng
\cite{yt2} proposed a stopping rule for terminating an ADT.
Chen \& Zheng \cite{cz} proposed an approach for degradation analysis
which makes inference directly on the lifetime distribution. Huang \& Dietrich \cite{hd},
used maximum likelihood estimation to estimate the model's parameters.
Joseph \& Yu \cite{jy}, developed a methodology for quality improvement
when degradation data are available as the response in the experiments.

ADT is usually very costly, since it requires destroying a
considerable number of products at each level of stress.
Moreover, determining suitable levels of stress for the experiment is not
straightforward. To handle these problems, step stress accelerated
degradation tests (SSADT) were proposed by Tseng \& Wen \cite{tw} in
a case study of LEDs. In a SSADT framework, each product is first
tested, subject to a pre-determined stress level for a specified
duration, and the degradation data are collected. A product which
survives until the end of the first step is again tested at a
higher stress level and for a different time duration. The stress
level is elevated step by step until an appropriate termination time
is reached. The advantage of the SSADT is that the number of test
units needed for conducting this test is relatively small.

To conduct a SSADT efficiently, special attention is paid to the
number of tested units (sample size), measurement frequency and
termination time. For ADT, Boulanger \& Escobar \cite{be} addressed
the problem of optimal determination of the stress levels as well as the
sample size for each stress level for a specified termination
time. The problem of optimizing the test design have been
extensively studied in recent years. Three commonly used
optimization criteria are the minimum approximated variance
\mbox{(Avar)} of the maximum likelihood estimators (MLE) of
reliability, mean time to failure (MTTF) and the quantiles of the
lifetime distribution. Tang \textit{et al.} \cite{tyx} discussed the
optimal planning of a simple SSADT and minimized the test cost while
achieving a requisite level of estimates, from which the optimal
number of inspections of the whole test can be determined. Liao \&
Tseng \cite{lt} used a Wiener process to model a typical SSADT
problem and then, under the additional constraint that the total
experimental cost does not exceed a predetermined budget, obtained
the optimal settings of the SSADT for minimizing the
\mbox{Avar} of the estimated 100$p^{\mbox{th}}$ percentile of the
product's lifetime distribution. They derived the optimal test plans
for the sample size, measurement frequency and termination time.
Tseng \textit{et al.} \cite{tbt} discussed the SSADT problem for a
gamma process and gave the optimal settings for minimizing the
\mbox{Avar} of the estimated MTTF of the lifetime distribution of
the product.

Recently, Pan \& Balakrishnan \cite{pb} proposed a step-stress
accelerated degradation test (SSADT) plan in which the stress level
is elevated when the degradation value of a product crosses a
pre-specified value. In such a plan the times of elevating the stress levels are
random and vary from product to product. However, such a plan can be extended to a more economic plan in two ways.
First one can extend the model such that the times of stress level elevating in the new model be
identical for all products, which enable us to use only one chamber (oven) for testing all test units. Second,
one can conduct the plan such that it does not require continuous inspection of the degradation by the experimenter (or may be electronic sensors).
In the later we don't need to spend any money on electronic sensors and also we have decreased the cost of measurement.

To extend the model, we assume that the
measurements are made at the end of consecutive intervals of fixed
duration. The experimenter elevates the stress level as soon as a
new measurement of at least one of the test units exceeds the threshold value.

In this paper, we propose a SSADT model when the degradation path
follows a gamma process. The measurements are made at times $kf$,
for $k=1,\ldots,M$, where $f$ is the measurement frequency in unit
of time and $M$ is the number of measurements. The stress level is
elevated as soon as a new measurement of the degradation of one at least of the test units exceeds the
threshold value. Next, under the constraint that the total
experimental cost does not exceed a pre-specified budget, the
optimal settings including the optimal threshold value, sample size,
measurement frequency and termination time are obtained by
minimizing the Avar of an estimated quantile of the
lifetime distribution of the product.

The paper is organized as follows. The model is described in Section II.
An optimization algorithm is presented in Section III to obtain the optimal
settings under budget constraints. Section IV presents a real data example
for illustrating the proposed algorithm and some concluding remarks are made in Section V.

\section{Model description}

Suppose that $n$ test units are subject to a degradation test. Let
$L(t|S)$ denote the degradation path of the product under a level
of stress $S$ at time $t$. Denote the use stress level (under the normal conditions) by $S_0$. The lifetime of
the $i^{\mbox{th}}$ unit under the normal conditions, $T_0^{(i)}$,
is defined the first time that $L(t|S_0)$ crosses the
critical value $D$, that is
\begin{equation}\label{lft}
T_0^{(i)}=inf\{t|L^{(i)}(t|S_0)\geq D\},
\end{equation}
in which $L^{(i)}(t|S_0),\; i=1,\ldots,n$, denotes the degradation
value of the $i^{\mbox{th}}$ unit under the stress $S_0$ at time $t$.

Pan \& Balakrishnan \cite{pb}, proposed a SSADT model in which the
$j^{\mbox{th}}$ stress level of the $i^{\mbox{th}}$ unit,
$i=1,\ldots,n$, $j=1,\ldots,m$, is elevated as soon as the
degradation of the unit passes the threshold value $\omega_j$
($\omega_0=0<\omega_1<\cdots<\omega_m<D$). Hence, the testing stress of the
$i^{\mbox{th}}$ unit, $i=1,\ldots,n$, under such a SSADT model can
be expressed for the stress values $S_0<S_1<\cdots<S_m$ as follows
\begin{eqnarray}\label{si}
\hspace{.1in}S^{(i)}= \left\{
\begin{array}{l c}
S_1& \tau_{i,0}=0\leq t<\tau_{i,1}\\
S_2& \tau_{i,1}\leq t< \tau_{i,2},\\
\vdots&\\
S_m & \tau_{i,m-1} \leq t< T,
\end{array}\right.
\end{eqnarray}
where $T$ is the termination time of the test and
\begin{equation}\label{tau}
\tau_{i,j}=\inf\{t|L^{(i)}(t|S_j)\geq \omega_j\},\quad
j=1,\ldots,m-1.
\end{equation}
in which $L^{(i)}(t|S_j),\; i=1,\ldots,n,\; j=1,\ldots,m$, denotes
the degradation value of the $i^{\mbox{th}}$ unit under the stress
levels $S_j$ at time $t$. We should keep in mind that the degradation values do not depend on the random variables $\tau_{i,j},\quad j=1,\ldots,m-1,$ and the random variables $\tau_{i,j},\quad j=1,\ldots,m-1,$ depend on the degradation values. In other words, we have
\begin{eqnarray}\label{si2}
\hspace{.1in}S^{(i)}= \left\{
\begin{array}{l c}
S_1& -\infty \leq L^{(i)}(t|S_1)\leq \omega_1,\\
S_2&  \omega_1\leq L^{(i)}(t|S_2)\leq \omega_2,\\
\vdots&\\
S_m & \omega_{m-1}\leq L^{(i)}(t|S_m)\leq D \quad \& \quad t<T.
\end{array}\right.
\end{eqnarray}
Also, it is worth noting the difference of notations in this paper with the paper of Pan and Balakrishnan \cite{pb}. In this paper, $L^{(i)}(t|S_j)$ denotes the cumulative degradation value of $i^{\rm th}$ item during the test, under the $j^{\rm th}$ stress level, while in Pan and Balakrishnan \cite{pb}, it denotes the cumulative degradation value of $i^{\rm th}$ item during the $j^{\rm th}$ stress level. 

They considered both Wiener and Gamma processes for the degradation path of the products.
In the following, we
assume that the independent increments of the degradation path of
the products follows a gamma process (Lawless \& Crowder \cite{lc});
that is, for fixed $t$ and $\Delta t$,
\begin{equation}\label{dl}
\Delta L(t|S_j)=L(t+\Delta t|S_j)-L(t|S_j)\sim Ga(\alpha_j\Delta
t,\beta),
\end{equation}
where $Ga(\alpha_j\Delta t,\beta)$
stands for the gamma distribution with shape and scale parameter
$\alpha_j\Delta t$ and $\beta$, respectively. Thus, we have, for $t>0$
\[L^{(i)}(t|S_j)\sim Ga\left(\alpha_j(t-\tau_{i,j})+\sum_{k=1}^{j-1}\alpha_i(\tau_{i,k}-\tau_{i,k-1})\right).\]

Park and Padgett \cite{pp1} provided the approximate
distribution of iid random differences $\tau_{i,k}-\tau_{i,k-1}$, for $k=1,\ldots,m-1$ as Birnbaum-–Saunders distribution with cdf
\begin{equation}\label{BS}
G_{\tau_{i,k}-\tau_{i,k-1}}(t)\approx\Phi\left(\frac{1}{\delta_1}
 \left(\sqrt{\frac{t}{\gamma_1}}-\sqrt{\frac{\gamma_1}{t}}\right)\right).
\end{equation}

Under the above settings and using \eqref{BS}, Pan and Balakrishnan \cite{pb} derived the approximate joint pdf of
$\tau_{i,1},\ldots,\tau_{i,m-1}$ as 
\begin{align}
g_{\tau_{i,1},\ldots,\tau_{i,m-1}}(t_1,\ldots,t_{m-1})&\approx \prod_{k=1}^{m-1}
\left\{\frac{\alpha_k}{2\sqrt{2\pi\beta(\omega_k-\omega_{k-1})}}\left[\left(\frac{\alpha_k(t_k-t_{k-1})}{\beta(\omega_k-\omega_{k-1})}\right)^{-1/2}\right.\right.\nonumber\\
&\left.\left.+\left(\frac{\alpha_k(t_k-t_{k-1})}{\beta(\omega_k-\omega_{k-1})}\right)^{-3/2}\right]\cdot\exp\left[-\frac{\beta(\omega_k-\omega_{k-1})}{2}\right.\right.\nonumber\\
&\left.\left.\quad\times\left(\frac{\alpha_k(t_k-t_{k-1})}{\beta(\omega_k-\omega_{k-1})}+
\frac{\beta(\omega_k-\omega_{k-1})}{\alpha_k(t_k-t_{k-1})}-2\right)\right]\right\}.\label{panbal}
\end{align}

The idea of using random time for elevating the stress level of SSADT is interesting.
However, implementing the above SSADT model with stress model (\ref{si}) is not economic, since the
main results for conducting a SSADT is that, we can economically use only one chamber (oven), which allows
us to elevate gradually the stress levels at specified (or random) times for all testing units. From equation (\ref{si}),
it is clear that $n$ chambers should be used simultaneously, since with current metrology technology it is not possible
to conduct such an SSADT plan within a single chamber.

A more economic model with random stress-elevating times can be conducted with the common stress
\begin{eqnarray}\label{s1}
\hspace{.1in}S= \left\{
\begin{array}{l c}
S_1& 0\leq t<\tau_{(1),1}\\
S_2& \tau_{(1),1}\leq t< \tau_{(1),2},\\
\vdots&\\
S_m & \tau_{(1),m-1} \leq t< T,
\end{array}\right.
\end{eqnarray}
for all test units, in which
\begin{equation}\label{tau1}
\tau_{(1),j}=\min\{\tau_{i,j};\;i=1,\ldots,n\},\quad
j=1,\ldots,m-1.
\end{equation}

Under the SSADT plan with stress model (\ref{s1}), the $j^{\mbox{th}}$ stress level is elevated
for all units as soon as the degradation of at least one of the units exceeds the threshold value $w_j$.
The test terminated at time $T$.

Using \eqref{panbal}, one can obtain the joint distribution of $\tau_{(1),1},\ldots,\tau_{(1),m-1}$, which is a rather complicated formula.

For the case $m=2$, since $\tau_{1,1},\ldots,\tau_{n,1}$ are independent and identically distributed, using \eqref{BS} $\tau_{(1),1}$ has the approximate cdf 
\begin{equation}\label{delmin}
 G_{\tau_{(1),1}}(t;\theta)\simeq 1-\left(1-\Phi\left(\frac{1}{\delta_1}
 \left(\sqrt{\frac{t}{\gamma_1}}-\sqrt{\frac{\gamma_1}{t}}\right)\right)\right)^n.\quad
\end{equation}

Also, for the case $m=3$, using \eqref{BS}, the approximate joint survival function of $\tau_{(1),1}$ and $\tau_{(1),2}$ is obtained as follows
\[{\rm Pr}(\tau_{(1),1}>t_1,\tau_{(1),2}>t_2;\theta)=\left({\rm Pr}(\tau_{1,1}>t_1,\tau_{1,2}>t_2;\theta)\right)^n,\]
where
\begin{align}
{\rm Pr}(\tau_{1,1}>t_1,\tau_{1,2}>t_2;\theta)&={\rm Pr}(\tau_{1,1}>t_1;\theta)-{\rm Pr}(\tau_{1,1}>t_1,\tau_{1,2}\leq t_2;\theta)\nonumber\\
&={\rm Pr}(\tau_{1,1}>t_1;\theta)(1-{\rm Pr}(\tau_{1,2}-\tau_{1,1}\leq t_2-t_1;\theta))\nonumber\\
&\simeq \left(1-\Phi\left(\frac{1}{\delta_1}
 \left(\sqrt{\frac{t_1}{\gamma_1}}-\sqrt{\frac{\gamma_1}{t_1}}\right)\right)\right)\nonumber\\
&\quad\times \Phi\left(\frac{1}{\delta_1}
 \left(\sqrt{\frac{t_2-t_1}{\gamma_1}}-\sqrt{\frac{\gamma_1}{t_2-t_1}}\right)\right).\quad\label{delmin2}
\end{align}

It is easy but a bit tedious to extend the above discussion for $m\geq 4$. For instanse, for $m=4$, we have 
\begin{align*}
{\rm Pr}(\tau_{1,1}>t_1,\tau_{1,2}>t_2,\tau_{1,3}>t_3;\theta)&={\rm Pr}(\tau_{1,1}>t_1;\theta)-{\rm Pr}(\tau_{1,1}>t_1,\tau_{1,2}\leq t_2;\theta)\\
&\quad-{\rm Pr}(\tau_{1,2}>t_2,\tau_{1,3}\leq t_3;\theta)\\
&\quad+
{\rm Pr}(\tau_{1,1}\leq t_1,\tau_{1,2}>t_2,\tau_{1,3}\leq t_3;\theta)\\
&={\rm Pr}(\tau_{1,1}>t_1;\theta){\rm Pr}(\tau_{1,2}-\tau_{1,1}\leq t_2-t_1;\theta)\\
&\quad-{\rm Pr}(\tau_{1,2}>t_2;\theta){\rm Pr}(\tau_{1,3}-\tau_{1,2}\leq t_3-t_2;\theta)\\
&\quad+
{\rm Pr}(\tau_{1,1}\leq t_1;\theta){\rm Pr}(\tau_{1,2}-\tau_{1,1}>t_2-t_1;\theta)\\
&\quad\times{\rm Pr}(\tau_{1,3}-\tau_{1,2}\leq t_3-t_2;\theta).
\end{align*}

The second step for conducting a more economic plan is to extend the model
with test stress model (\ref{s1}) such that it does not require continuous inspection of the
degradation level of test units by the experimenter or electronic sensors. In the following, we
describe the extended model with discrete inspection times.

Suppose we want to make $M$ measurements for each
unit with a measurement frequency per $f$ unit of time.
Note that for the stress model (\ref{s1}), since the stress elevating time is random, we can only determine
the total number of measurements and the number of measurements under each stress level is random.
Under the $j^{\mbox{th}}$ stress
level, $j=1,\ldots,m$, we elevate the stress level, as soon as a new measurement
of the degradation of at least one of the test units exceeds the threshold value
$\omega_j$. The testing stress of the $i^{\mbox{th}}$ unit,
$i=1,\ldots,n$, under such a model can be expressed as follows
\begin{eqnarray}
\hspace{.1in}S= \left\{
\begin{array}{l c}
S_1& 0\leq t<\kappa_{1} f\\
S_2& \kappa_{1} f \leq t< \kappa_{2} f,\\
\vdots&\\
S_m & \kappa_{m-1} f \leq t< Mf,
\end{array}\right.\label{model}
\end{eqnarray}
where
\begin{equation}\label{kap}
\kappa_{j}=\min\{M, \left[\left[\frac{\tau_{(1),j}}{f}\right]\right]\},\quad
j=1,\ldots,m-1,
\end{equation}
in which $[[x]]$ stands for the smallest integer greater than $x$ and $\tau_{(1),j}$ is
defined in (\ref{tau1}).

The random variables $\kappa_{j}$ are discrete with the obvious property that
$1\leq\kappa_{1}\leq\kappa_{2}\leq\cdots\leq\kappa_{m-1}\leq M$, with probability 1.
Also, it is straightforward that, if for some $j^*$,
$\kappa_{j^*}=M$, we have $\kappa_{j}=M$, for all $j>j^*$
and $\min\{T_{(1),m}/f,M\}=M$. Therefore, if for some $j^*$,
$\kappa_{j^*}=M$, the testing stress takes only the values $S_1<\cdots<S_{j^*}$
during the total time interval, $[0,\;Mf]$.

For the case $m=2$, we have
\begin{align}
{\rm Pr}(\kappa_{1}=k;{\theta})&={\rm Pr}((k-1)f<\tau_{(1),1}\leq kf;{\theta})\nonumber\\
&\approx G_{\tau_{(1),1}}(kf;\theta)-G_{\tau_{(1),1}}((k-1)f;\theta),\quad k=1,\ldots,M-1,\label{pmfk}
\end{align}
and
\begin{equation}\label{pmfk2}
{\rm Pr}(\kappa_{1}=M;{\theta})={\rm Pr}(\tau_{(1),j}> (M-1)f;{\theta})\approx 1-G_{\tau_{(1),1}}((M-1)f;\theta),
\end{equation}
where $G_{\tau_{(1),1}}(t;\theta)$ is given in \eqref{delmin}.

Also, for the case $m=3$, the joint pmf of $\kappa_{1}$ and $\kappa_{2}$ is
\begin{align}
{\rm Pr}(\kappa_{1}=k_1,\kappa_{2}=k_2;{\theta})&=P((k_1-1)f<\tau_{(1),1}\leq k_1f, (k_2-1)f<\tau_{(1),2}\leq k_2f;{\theta}),\nonumber\\
& k_1,k_2=1,\ldots,M-1,\label{pmfk3}
\end{align}
and simillar expressions are derived easily for ${\rm Pr}(\kappa_{1}=M,\kappa_{1}=k_2;{\theta})$, ${\rm Pr}(\kappa_{1}=k_1,\kappa_{1}=M;{\theta})$ and ${\rm Pr}(\kappa_{1}=M,\kappa_{1}=M;{\theta})$, which are all approximated 
using the joint survival function given in \eqref{delmin2}. Similar formulas can also be obtained for the joint pmf of $\kappa_{1},\ldots,\kappa_{m-1}$, for $m\geq 4$. 

Suppose that the relation between $\alpha_j$ and the stress level $S_j$
is modeled by the Arrhenius reaction rate model  as
\[\alpha_j=exp (a+\frac{b}{273+S_j}),\; j=0,1,\ldots,m,\]
for two unknown parameters $a$ and $b$.

For $k=0,\ldots,M-1$ and $i=1,\ldots,n$, let $G_{ki}$ denote the degradation increment of the $i^{\mbox{th}}$ unit at time $(k+1)f$ relative
to time $kf$, that is
\[G_{ki}=L^{(i)}((k+1)f)-L^{(i)}(kf),\]
where $L^{(i)}(kf)$ is the level of degradation of the $i^{\rm th}$ unit at time $kf$.
We have $L^{(i)}(t)=L^{(i)}(t|S_j)$ for all $t\in[kf,(k+1)f]$ for $k=\kappa_{j-1},\ldots,\kappa_{j}-1$ with $\kappa_{0}=0$, $\kappa_{m}=M$, and $j=1,\ldots,m$.

Using (\ref{dl}), the conditional distribution of $G_{ki}$ given $\kappa_{1}$, for $k=\kappa_{j-1},\ldots,\kappa_{j}-1$, is $Ga(\alpha_jf,\beta)$, $j=1,\ldots,m$, in which we assume $\kappa_0=0$ and $\kappa_m=M$. Therefore, given $G_{ki}=g_{ki}$ and $(\kappa_1=k_1,\ldots,\kappa_{m-1}=k_{m-1})$, the likelihood function of $\theta=(a,b,\beta)$ is given by
\begin{equation}\label{lt}
L(\theta)={\rm P}_{\theta}(\kappa_1=k_1,\ldots,\kappa_{m-1}=k_{m-1})\prod_{i=1}^n\prod_{j=1}^m\prod_{k=k_{j-1}}^{k_{j}-1}
\frac{g_{ki}^{f\alpha_{j}-1}e^{-g_{ki}/\beta}}{\Gamma(f\alpha_{j})\beta^{f\alpha_{j}}},
\end{equation}
where ${\rm P}_{\theta}(\kappa_1=k_1,\ldots,\kappa_{m-1}=k_{m-1})$ is given in
(\ref{pmfk}) to (\ref{pmfk3}) for $m=2,3$ and  similar formulas can also be obtained for the joint pmf of $\kappa_{1},\ldots,\kappa_{m-1}$, for $m\geq 4$. 

\section{The Optimal Design}

We develope the optimization algorithm for the case $m=2$. The generalization of the proposed algorithm for larger values of 
$m$ is straightforward. For the case $m=2$, the optimization of the SSADT model in (\ref{model}) consists of
finding the optimal values of $n$, $M$, $f$, and the optimal value of
$\omega_1$. We concern the problem of
minimizing the \mbox{Avar} of an estimate of a quantile of the
product's lifetime distribution as an optimization criterion.
The approximated cdf of the lifetime of the product
under the use stress level, $S_0$, which is defined as in
(\ref{lft}), is (see Park and Padgett, \cite{pp1})
\begin{equation}\label{ef}
F_0(t)\simeq \Phi\left(\frac{1}{\alpha_0^*}
\left(\sqrt{\frac{t}{\beta_0^*}}-\sqrt{\frac{\beta_0^*}{t}}\right)\right),
\end{equation}
where
\[\alpha_0^*=\sqrt{\frac{\beta}{D}},\quad \mbox{and} \quad \beta_0^*=\frac{D}{\beta\alpha_0}.\]

Using (\ref{ef}), one can obtain the 100$p^{\mbox{th}}$ percentile of $T_0$ as
$\xi_p=F_0^{-1}(p), \; 0<p<1$. The MLE of $\xi_p$,
$\hat{\xi_p}=\hat{F}^{-1}_0(p)$, is obtained by substituting the
MLEs of $\beta$ and $\alpha_0$ into $F_0^{-1}(p)$. These values can
be obtain by maximizing the likelihood function of
$\theta=(a,b,\beta)$ in (\ref{lt}).

The \mbox{Avar} of $\hat{\xi}_p$ can be obtained
as a function of $\omega_1$, based on \mbox{Avar} of the MLE (The
inverse of the Fisher information matrix). Using the delta method
we have
\[\mbox{Avar}(\hat{\xi}_p;{\omega_1},n,f,M)=\frac{1}{(\hat{f}_0(\hat{\xi_p}))^2}h^TI^{-1}(\hat{\theta}(\omega_1))h,\]
where $f_0$ is the corresponding pdf of the cdf  in (\ref{ef}), the transpose of vector $h$ is
$$h^T=\left(\frac{\partial F_0(\hat{\xi_p})}{\partial
a},\frac{\partial F_0(\hat{\xi_p})}{\partial b},\frac{\partial
F_0(\hat{\xi_p})}{\partial\beta}\right),$$
and $I(\theta)$ is the
Fisher information matrix of the likelihood in (\ref{lt})
which is calculated and given in Appendix and
$\hat{\theta}(\omega_1)$ is the MLE of $\theta$ for a fixed
$\omega_1$.

We have
 \[\frac{\partial F_0(t)}{\partial a}=\frac{1}{2\alpha_0^*}\left(\sqrt{\frac{t}{\beta_0^*}}+\sqrt{\frac{\beta_0^*}{t}}\right)
 \phi_0(t),\]

 \[\frac{\partial F_0(t)}{\partial b}=\frac{1}{273+S_j}\frac{\partial F_0(t)}{\partial a},\]

\[\frac{\partial F_0(t)}{\partial \beta}=\frac{1}{\beta\alpha_0^*}\sqrt{\frac{\beta_0^*}{t}} \phi_0(t),\]
 and
 \[f_0(t)=\frac{1}{2\alpha_0^*\sqrt{t\beta_0^*}}(1+\frac{\beta_0^*}{t})\phi_0(t),\]
  where
  $\phi_0(t)=\phi\left(\frac{1}{\alpha_0^*}\left(\sqrt{\frac{t}{\beta_0^*}}-\sqrt{\frac{\beta_0^*}{t}}\right)\right)$
  and $\phi$ stands for the pdf of the standard normal
  distribution.\\

 The total cost of the experiment $TC(n,f,M)$
 is given by
 \[TC(n,f,M)=C_{op}\cdot f\cdot M+C_{mea}\cdot n\cdot M+C_{it}\cdot n,\]
 where $C_{op}$ is the unit cost of operation per unit of time, $C_{mea}$ is the
 unit cost of measurement, and $C_{it}$ is the unit cost of items.
The optimization criterion is to find the values
$n^*,f^*,M^*,\omega_1^*$, which minimize
$\mbox{Avar}(\hat{\xi}_p;{\omega_1},n,f,M)$, subject to $TC(n^*,f^*,M^*)\leq
C_b$, where $C_b$ is the total budget for conducting the degradation
experiment. 

Since the parameter spaces ${\cal N}=\{(n,f,M): TC(n,f,M)\leq
C_b\}$ and $\Omega=\{\omega_1: \omega_1\in(0,D)\}$ are independent we have
\begin{eqnarray}\label{infinf}
\inf_{(n,f,M,\omega_1)\in{\cal N}\times\Omega}\mbox{Avar}(\hat{\xi}_p;{\omega_1},n,f,M)=\inf_{\omega_1\in\Omega}\inf_{(n,f,M)\in{\cal N}}\mbox{Avar}(\hat{\xi}_p;{\omega_1},n,f,M).
\end{eqnarray}
Thus, although there is no analytic expression for the solution of the
optimization problem of finding $\inf_{(n,f,M)\in{\cal N}}\mbox{Avar}(\hat{\xi}_p;{\omega_1},n,f,M)$, due to the integer restriction on the optimality parameters $n$, $f$ and $M$, 
 the global and unique minimum of $\mbox{Avar}(\hat{\xi}_p;{\omega_1},n,f,M)$ can be found by 
searching the minimum through all possible values of $n$, $f$ and $M$ (see for example Tseng \textit{et al.} \cite{tbt}), for a fixed value of $\omega_1$ and then applying the common minimization algorithms for minimization of the 
continuous function $\min_{(n,f,M)}(\mbox{Avar}(\hat{\xi}_p;{\omega_1},n,f,M))$ of $\omega_1$. 

Summing up, the optimal solution of the above optimization problem can be
determined by the Algorithm I below.
\vspace{0.5cm}
\begin{algorithm}

Step 1) Define the function $\varphi(\omega_1)$ with domain
$(0,D)$ as follows:
\begin{description}
\item Step 1-1) Compute the largest possible number for $n$, when
$f=1$ and  $M=2$ (one measurement for each stress level), which is
equal to
\[n_{max}=\left[\frac{C_b-2C_{op}}{2C_{mea}+C_{it}}\right].\]
\item Step 1-2) Set $n=1$.
\item Step 1-3) Compute the largest possible number for $f$, for the fixed $n$, and $M=2$, which is
\[f_{max}=\left[\frac{C_b-2C_{mea}.n-n C_{it}}{2C_{op}}\right].\]
\item Step 1-4) Set $f=1$.
\item Step 1-5) Let $M=\left[\frac{C_b-nC_{it}}{nC_{mea}+fC_{op}}\right]$.
\item Step 1-6) Compute $\mbox{Avar}(\hat{\xi}_p;{\omega_1},n,f,M)$.
\item Step 1-7) Set $f=f+1$, and repeat steps 1-5 and 1-6 until $f=f_{max}$.
\item Step 1-8) Set $n=n+1$ and repeat steps 1-3 through 1-7 until $n=n_{max}$.
\item Step 1-9) Let $(n^*(\omega_1),f^*(\omega_1),M^*(\omega_1))=\arg\min_{(n,f,M)}(\mbox{Avar}(\hat{\xi}_p;{\omega_1},n,f,M))$.
\item Step 1-10) Return
$$\varphi(\omega_1)=\min_{(n,f,M)}(\mbox{Avar}(\hat{\xi}_p;{\omega_1},n,f,M)).$$
\end{description}
 Step 2) Let $\omega_1^*=\arg \inf_{\omega_1}[\varphi(\omega_1)]$
and $(n^*,F^*,M^*)=(n^*(\omega_1^*),f^*(\omega_1^*),M^*(\omega_1^*))$.
\end{algorithm}
It is worth noting that the above optimization algorithm can easily be extended for larger values of $m$. For example generalization to  the case $m=3$, can be easily made by replacing 
$\varphi(\omega_1)$ and $(0,D)$ in Step 1 of Algorithm I with $\varphi(\omega_1,\omega_2)$ and $(0,D)^2$, respectively. 
\section{Numerical illustration}
\textbf{
\begin{table}
\centering \caption{Optimal SSADT plan for minimizing
$\mbox{Avar}(\hat{\xi}_{p};\omega_1,n,f,M)$ for different values of $p$, under the
budget constraint.\label{opt}}
\begin{tabular}{c | c c c c }
\hline\hline
$p$ & $\hat{\xi}_{p}$ & minimum C.V. & $\omega^{*}_1$ & $G_{\tau_{(1),1}}(T^*;\hat{\theta})$ \\
\hline
0.1 & 292795.6 & 0.1389 & 0.0507 & 0.9999950 \\
0.2 & 307152.4 & 0.1358 & 0.0505 & 0.9999951 \\
0.3 & 317950.3 & 0.1335 & 0.0504 & 0.9999952 \\
0.4 & 327481.6 & 0.1318 & 0.0503 & 0.9999953 \\
0.5 & 336650.3 & 0.1300 & 0.0502 & 0.9999954 \\
0.6 & 346075.7 & 0.1283 & 0.0501 & 0.9999955 \\
0.7 & 356450.0 & 0.1266 & 0.0500 & 0.9999956 \\
0.8 & 368981.1 & 0.1245 & 0.0499 & 0.9999957 \\
0.9 & 387073.5 & 0.1218 & 0.0497 & 0.9999959 \\
\hline
\end{tabular}
\end{table}
}
\begin{figure}
\centerline{\psfig{figure=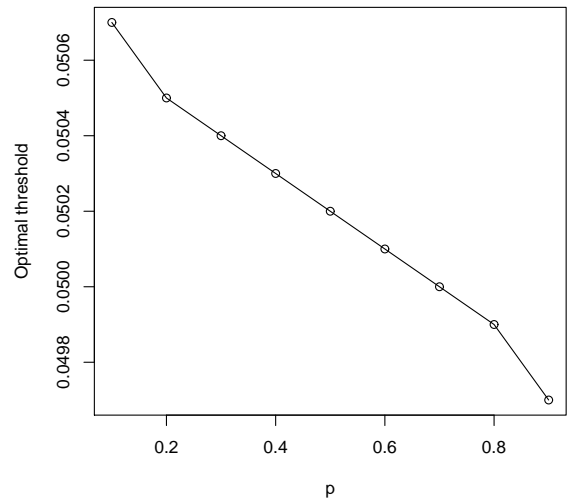,width=8cm}}
\caption{\label{thp} \small{Plot of $\omega^{*}_1$ versus $p$.}}
\end{figure}

\begin{figure}
\centerline{\psfig{figure=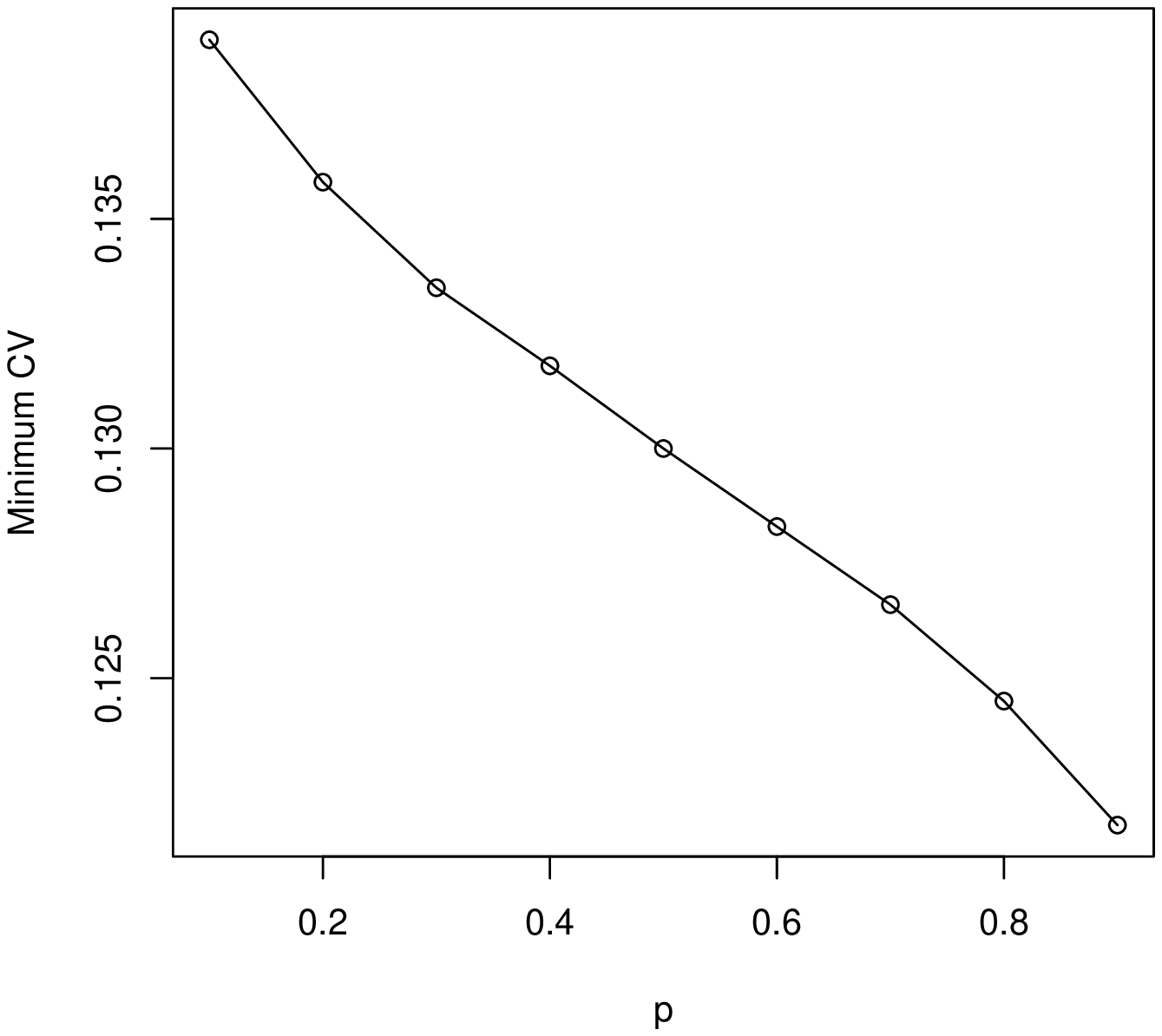,width=8cm}}
\caption{\label{cvp} \small{Plot of minimum C.V. as a function of
$p$.}}
\end{figure}

We consider the data from the carbon-film-resistor problem described
by Meeker \& Escobar \cite{me}, for the purpose of illustration of
the proposed procedure. Increasing the resistance value of the
carbon-films resistors over time reduces the performance quality of
the products and leads to failure. The failure occurs as soon as the
percent increase in resistance hits a threshold value $D=5$, under
the use operating temperature $S_0=50^{\circ}$C. Suppose that the
cost of operation, $C_{op}$ is {\$}1.9 per unit of time, the cost of
measurement, $C_{mea}$ is {\$}1.3 for each measurement, the cost of each item, $C_{it}$
is {\$}53, the unit time is 4 hours and the total budget is
{\$}1500.

For obtaining the optimal design we need some initial values for the parameters. These initial values are usually obtained from
a pilot study. The pilot study need not to be even a step stress test. Tseng \textit{et al.} \cite{tbt} used this data set
and obtained an initial estimate of the parameters as follows
\[(\hat{a},\hat{b},\hat{\beta})=(4.11,-4006.46,0.0594).\]
We use the above estimates as the true parameter vector of the SSADT
model with gamma degradation process and $S_1=83^{\circ}$C
and $S_2=133^{\circ}$C. Algorithm I is used with
software R.2.14 for the optimization process.
Interestingly, the optimal sampling setting is
obtained as
$$(n^*,f^*,M^*)=(13,52,7),$$
for all values of $p=0.1(0.1)0.9$. Hence, the optimal
total time of experiment is equal to $T^*=4fM=1456$. The optimal threshold values
$\omega_1^*$, are given in Table \ref{opt} which also includes
the estimated $\xi_p$, minimum C.V. (the minimum coefficient of variation of $\hat{\xi_p}$ equal
to $\sqrt{\varphi(\omega_1^*)}/\hat{\xi_p}$, where $\varphi$ is defined in Algorithm I)
and the probability of stress level elevation before the end of the
experiment, that is $G_{\tau_{(1),1}}(T^*;\hat{\theta}),$ for $p=0.1(0.1)0.9$. For example,
for estimating the median time to failure of the product ($p=0.5$),
the optimal setting is to elevate the stress level as soon as a new measurement
of the degradation of at least one of the test units exceeds the threshold value $\omega_1^*=0.0502$, which occurs
with probability 0.9999954.

Figure \ref{thp} shows the plot of
$\omega^{*}_1$ versus $p$. The optimal value of $\omega^{*}_1$
decreases as $p$ increases. The reason is that the degradation values under stress level $S_2$
provide more information about upper quantiles ($\xi_p$ for greater values of $p$) than those under $S_1$.

Also the minimum C.V. is plotted as a
function of $p$ in Figure \ref{cvp}. As we can see from Figure \ref{cvp}, the estimation precision
increases for upper quantiles. This means that the SSADT plan is more suitable for
estimation of the upper quantiles of the product's lifetime.

\section{Sensitivity and stability analysis}

\begin{table*}
\small\centering \caption{Stability analysis of parameter estimation for different values of $(n,f,M,\omega_1)$.\label{stab}}
\begin{tabular}{c | c c c c c c }
\hline\hline
$(n,f,M,\omega_1)$ & Bias($\hat{a}$) & MSE($\hat{a}$) & Bias($\hat{b}$) &  MSE($\hat{b}$) & Bias($\hat{\beta}$) & MSE($\hat{\beta}$)\\
\hline
(13, 52, 7, 0.0504)  & 0.125667 & 0.503172 & 0.460590 & 0.212146 & -0.001530 & 9.228e-05 \\
(13, 52, 7, 0.0502)  & 0.204601 & 0.912977 & 0.460766 & 0.212310 & -0.003645 & 9.973e-05\\
(13, 62, 6, 0.0818)  & 0.153269 & 0.647985 & 0.460649 & 0.212201 & -0.002000 & 0.000117 \\
(13, 62, 6, 0.0912)  & 0.133570 & 0.535452 & 0.460604&  0.212159 & -0.004489  & 0.000113\\
(13, 204, 2, 0.6181) & 0.188061 & 0.595449 & 0.460837&  0.212375 & -0.005635   &0.000303 \\
\hline
\hline
\end{tabular}
\end{table*}

\begin{table}
\centering \caption{Sensitivity analysis for $p=0.3,\; 0.5,\; 0.7$ and different values of $\epsilon_1$,
$\epsilon_2$ and $\epsilon_3$.\label{sens}}
\begin{tabular}{c c c | c c c c c c }

\hline\hline
 & &  &  & $p=0.3$ &  &  &  \\
\hline
$\epsilon_1$ & $\epsilon_2$ & $\epsilon_3$ & min.C.V. & $\omega_1^*$ & $n^*$ & $f^*$ & $M^*$ \\
\hline
0 & 0 & 0                & 0.1335 & 0.0504 & 13 & 52 & 7 \\%
48.5{\%}& 0 & -3.5{\%}    & 0.0519 & 0.1517 & 13 & 62 & 6 \\%
-7.9{\%}& 0 & 4.9{\%}     & 0.2959 & 0.6181 & 13 & 204 & 2\\%
0 & 1.9{\%} & -1.9{\%}    & 0.2797 & 0.6181 & 13 & 204 & 2 \\%
0 & -6.6{\%} & 4.3{\%}   & 0.0992 & 0.0818 & 13 & 62 & 6 \\%
-1.8{\%}& 1.8{\%} & 0    & 0.2887 & 0.6181 & 13 & 204 & 2 \\%
6.5{\%}& -6.5{\%} & 0    & 0.0877 & 0.0893 & 13 & 62 &  6\\%
2.8{\%}&2.8{\%}& -1.9{\%} &0.2778 & 0.6181 & 13 & 204 & 2 \\%
-3.9{\%}&-1.8{\%}& 5.8{\%}& 0.1224 & 0.0744 & 13 & 62 & 6 \\%
\hline
 & &  &  & $p=0.5$ &  &  &  &  \\
\hline
$\epsilon_1$ & $\epsilon_2$ & $\epsilon_3$ & min.C.V. & $\omega_1^*$ & $n^*$ & $f^*$ & $M^*$ \\
\hline
0 & 0 & 0                 & 0.1300 & 0.0502 & 13 & 52 & 7 \\%
41.5{\%}& 0 & -2.5{\%}    & 0.0519 & 0.1517 & 13 & 62 & 6 \\%
-7.8{\%}& 0 & 4.7{\%}     & 0.2868 & 0.6181 & 13 & 204 & 2\\%
0 & 1.8{\%} & -1.8{\%}    & 0.2704 & 0.6181 & 13 & 204 & 2 \\%
 0&-8.0{\%}  &8.0{\%}     & 0.0903 & 0.0912 & 13 & 62 &  6\\
 -1.6{\%} &1.6{\%} &0     & 0.2763 & 0.6181 & 13 & 204 & 2 \\
 7.2{\%}&-7.0{\%} &0      &0.08219  &0.0923 &13  &62  &6  \\
 2.7{\%}& 2.7{\%} &-2.2{\%}&0.2690&0.6181& 13 & 204 & 2 \\
-9.3{\%}&-9.4{\%} &9.3{\%} &0.1012  &0.0832  &13  &62  &6  \\
\hline
 & &  &  & $p=0.7$ &  &  &  &  \\
\hline
$\epsilon_1$ & $\epsilon_2$ & $\epsilon_3$ & min.C.V. & $\omega_1^*$ & $n^*$ & $f^*$ & $M^*$ \\
\hline
 0 &0 &0                 &0.1266 &0.0500 &13&52&7\\%
 75{\%}& 0 &-67{\%}      &0.0289 &0.1090 & 13 &62  &6  \\
 -31.5{\%}&0 & 31.5{\%}  &0.4516 &0.9999 &13  &204  &2  \\
 0&0.9{\%} &-7{\%}       &0.2500 &0.6181 &13  &204  &2  \\
  0&-9.9{\%} &9.0{\%}    &0.0802 &0.1020&13  &62  &6  \\
-1.5{\%}&1.5{\%} &0       &0.2669 &0.6181&13  &204  &2  \\
9.9{\%}&-9.9{\%} &0       &0.0660 &0.1164 &13  &62  &6  \\
2.7{\%} &2.7{\%} &-2.3{\%}&0.2616 &0.8181&13  &204  &2  \\
-0.12 &-0.12 &0.12       &0.0917 &0.0931&13  &62  &6  \\
\hline \hline
\end{tabular}
\end{table}

The optimal design depends on the initial estimates of the parameters. Hence, it is
appropriate for the optimization algorithm to be rather robust for departures from the real values of the parameters.
In order to check the effect of variation of $\hat{\theta}$ on the optimal setting
$(n^*,f^*,M^*,\omega_1^*)$, we consider a pilot run on the SSADT plan for various combinations of
$((1+\epsilon_1)a,(1+\epsilon_2)b,(1+\epsilon_3)\beta)$.
Table \ref{sens} presents some values of $\epsilon_1$,
$\epsilon_2$ and $\epsilon_3$ which cause a change in the
optimal setting $(n^*,f^*,M^*)$ and the corresponding $\omega_1^*$, for $p=0.3,\;0.5,\;0.7$.

From the values of Table \ref{sens}, it is clear that the optimization algorithm is quite robust for small departures from $\theta$, since
the changes in decision variables $(\omega_1^*,n^*,f^*,M^*)$ occur for relatively large values of $(\epsilon_1,\epsilon_2,\epsilon_3)$.
In fact, the variation of $\hat{\theta}$ affect on the value of $\omega_1^*$, and this leads to
a more robust structure for the remaining decision variables $(n^*,f^*,M^*)$.

To examine the stability of the optimal test plan, we perform a simulation
study with $\hat{\theta}=(4.11,-4006.46,0.0594)$ taken as the true parameter of
the SSADT model. A Monte Carlo simulation study with 10,000 iterations is performed
and the ML estimate of the $\theta$ is obtained for each iteration using the log-likelihood
function in (\ref{lt}), for some values of the decision vector $(n,f,M,\omega_1)$.
The bias and the mean square error (MSE) of the estimates are obtained and tabulated in Table \ref{stab}.
From the values of Table \ref{stab} it can be observed that the estimates of the parameters are quite stable,
for small departures from the optimal plan.

\section{Conclusion}
This paper presents an approach for performing an SSADT experiment
in which the stress level is elevated on the basis of the degradation value.
In the proposed method, there is no need to observe the value of the degradation
continually. The stress level is elevated as soon as a new measurement of the degradation
exceeds the threshold value. We consider this threshold value as a decision variable in the
optimization problem of the plan as well as the usual variables such as sample size, measurement frequency and
number of measurements. An algorithm is proposed for minimizing the approximated variance of the ML estimator of
the $p^{\mbox{th}}$ quantile of the lifetime of the products, under the budget constraints. It is observed that
the proposed optimization algorithm is quite robust for departures from the values of the parameters.
Also the estimates of the parameters are quite stable for small departures from the optimized plan.

\section*{Appendix [The Fisher information matrix]}
Since, the likelihood function in \eqref{lt} satisfies the regularity conditions, specially that the paramater space does not depend on the observations, the Fisher information matrix of the SSADT data about the parameter $\theta$ is as follows
\[I(\theta)=-\rm{E}
\left[
\begin{array}{c c c}
\frac{\partial^2}{\partial a^2}l(\theta) & \frac{\partial^2}{\partial a\partial b}l(\theta) &\frac{\partial^2}{\partial a\partial\beta}l(\theta)\\
\frac{\partial^2}{\partial a\partial b}l(\theta) & \frac{\partial^2}{\partial b^2}l(\theta)&\frac{\partial^2}{\partial b\partial\beta}l(\theta)\\
\frac{\partial^2}{\partial a\partial\beta}l(\theta)& \frac{\partial^2}{\partial b\partial \beta}l(\theta) &\frac{\partial^2}{\partial \beta^2}l(\theta)
\end{array}\right],
\]
where, $l(\theta)$ is the log-likelihood of $\theta$ which, for the case $m=2$ of (\ref{lt}), is
\begin{align*}
l(\theta)&=\log {\cal{P}}_{\kappa_1}(\kappa_1)+\sum_{i=1}^n\sum_{j=1}^2\sum_{k=\kappa_{j-1}}^{\kappa_{j}-1}\left\{(f\alpha_{j}-1)\log
g_{ki}-g_{ki}\beta -\log \Gamma(f\alpha_{j})-f\alpha_{j}\log
\beta\right\},\\
\end{align*}
in which $\kappa_{0}=0$ and $\kappa_{2}=M$.

For the sake of brivity we define the following notations:
$$c_1(t)=\frac{1}{\delta_1}\left(\sqrt{\frac{t}{\gamma_1}}-\sqrt{\frac{\gamma_1}{t}}\right),$$
$$c_2(t)=\frac{1}{\delta_1}\left(\sqrt{\frac{t}{\gamma_1}}+\sqrt{\frac{\gamma_1}{t}}\right),$$
$$\Delta [\varphi(kf)]=\varphi(kf)-\varphi((k-1)f),$$
for any function $\varphi$, and
$$E_{\kappa}=\sum_{k=1}^{M-1}k\mbox{P}(\kappa_i=k).$$

Also, let $$g_1(t)=n[1-\Phi(c_1(t))]^{n-1}\phi(c_1(t))$$ and
\begin{align*}
g'_1(t)=-n(n-1)[1-\Phi(c_1(t))]^{n-2}\phi^2(c_1(t))-nc_1(t)[1-\Phi(c_1(t))]^{n-1}\phi(c_1(t))
\end{align*}
denote the differentiation of  $G_{\tau_{(1),1}}(t)$ and $g_1(t)$ with respect to $c_1(t)$, respectively.

Since we
have for $j=1,2$ and $i=1,\ldots,n$
\begin{eqnarray*}
E(\log g_{ki}|\kappa_{i,1})&=&\sum_{j=1}^2\psi_0(f{\alpha_{j}})I_{\{\kappa_{j-1},\ldots,\kappa_{j}-1\}}(k)
 I_{1,\ldots,M-1}(\kappa_{1})+\psi_0(f{\alpha_{1}}) I_{M}(\kappa_{1})+\log \beta,
\end{eqnarray*}
we obtain
\begin{eqnarray*}
E \left[\frac{-\partial^2}{\partial a^2}l(\theta)\right]\!&\!=\!&\!
nf^2\left\{[\alpha_1^2\psi_1(f\alpha_1)-\alpha_2^2\psi_1(f\alpha_2)].E_{\kappa}\right.\\&&+M\alpha_1^2\psi_1(f\alpha_1)
[1-G_{\tau_{(1),1}}((M-1)f;\theta)]
\\&&\left.+M\alpha_2^2\psi_1(f\alpha_2)G_{\tau_{(1),1}}((M-1)f;\theta)\right\}+\frac{A}{4},
\end{eqnarray*}
where $\psi_0(t)=\frac{d}{dt}\log \Gamma(t)$ and $\psi_1(t)=\frac{d^2}{dt^2}\log \Gamma(t)$ are the digamma and trigamma functions, respectively, and
\begin{align*}
A&=\sum_{k=2}^{M-1}\frac{\left[\Delta[c_2(kf)g_1(kf)]\right]^2}{\Delta[G_{\tau_{(1),1}}(kf;\theta)]}
+\frac{\left[c_2(f)g_1(f)\right]^2}{G_{\tau_{(1),1}}(f;\theta)}+\frac{\left[c_2((M-1)f)g_1((M-1)f)\right]^2}{1-G_{\tau_{(1),1}}((M-1)f;\theta)}\\
&\quad-\sum_{k=2}^{M-1}\left[\Delta[c_1(kf)g_1(kf)+(c_2(kf))^2g'_1(kf)]\right]-\left[c_1(f)g_1(f)+(c_2(f))^2g'_1(f)\right]\\
&\quad+\left[c_1((M-1)f)g_1((M-1)f)\right.\left.+(c_2((M-1)f))^2g'_1((M-1)f)\right].\\
\end{align*}
Similarly, we have
\begin{align*}
E \left[\frac{-\partial^2}{\partial b^2}l(\theta)\right]&=
nf^2\left\{\left[\left(\frac{\alpha_1}{273+S_1}\right)^2\psi_1(f\alpha_1)
-\left(\frac{\alpha_2}{273+S_2}\right)^2\psi_1(f\alpha_2)\right].E_{\kappa}\right.\\
&\quad\quad\quad\quad+M\left(\frac{\alpha_1}{273+S_1}\right)^2\psi_1(f\alpha_1)[1-G_{\tau_{(1),1}}((M-1)f;\theta)]\\
&\quad\quad\quad\quad+M\left.\left(\frac{\alpha_2}{273+S_2}\right)^2\psi_1(f\alpha_2)G_{\tau_{(1),1}}((M-1)f;\theta)\right\}\\
&\quad+\left(\frac{1}{273+S_1}\right)^2\frac{A}{4}
\end{align*}
and
\begin{align*}
E \left[\frac{-\partial^2}{\partial a\partial b}l(\theta)\right]&=nf^2\left\{\left[\left(\frac{\alpha_1^2}{273+S_1}\right)\psi_1(f\alpha_1)
-\left(\frac{\alpha_2^2}{273+S_2}\right)\psi_1(f\alpha_2)\right].E_{\kappa}\right.\\
&\quad\quad\quad\quad+M\left(\frac{\alpha_1^2}{273+S_1}\right)\psi_1(f\alpha_1)[1-G_{\tau_{(1),1}}((M-1)f;\theta)]\\
&\quad\quad\quad\quad+M\left.\left(\frac{\alpha_2^2}{273+S_2}\right)\psi_1(f\alpha_2)G_{\tau_{(1),1}}((M-1)f;\theta)\right\}\\
&\quad+\left(\frac{1}{273+S_1}\right)\frac{A}{4}.
\end{align*}
Since
\begin{eqnarray*}
E(g_{ki}|\kappa_{i,1})&=&\sum_{j=1}^2f\alpha_j\beta I_{\{\kappa_{j-1},\ldots,\kappa_{j}-1\}}(k)
I_{1,\ldots,M-1}(\kappa_{1})+f\alpha_1\beta I_{M}(\kappa_{1})+\log \beta,
\end{eqnarray*}
we have similarly
\begin{align*}
E \left[\frac{-\partial^2}{\partial\beta
\partial a}l(\theta)\right]&=\frac{\;C\gamma_1}{2\beta}+\frac{nf}{\beta}\left\{[\alpha_1-\alpha_2].E_{\kappa}+M\alpha_1[1-G_{\tau_{(1),1}}((M-1)f;\theta)]\right.\\
&\left.+M\alpha_2G_{\tau_{(1),1}}((M-1)f;\theta)\right\},
\end{align*}

where
\begin{align*}
C &=\sum_{k=2}^{M-1}\frac{\Delta[\sqrt{\frac{\alpha_1}{kf}}g_1(kf)]
\Delta[c_2(kf)g_1(kf)]}{\Delta[G_{\tau_{(1),1}}(kf;\theta)]}-\sum_{k=2}^M\Delta\left[\sqrt{\frac{\alpha_1}{kf}}g_1(kf)+\gamma_1{\frac{\alpha_1}{kf}}g'_1(kf)\right]\\
&+\frac{\sqrt{\frac{\alpha_1}{f}}g_1(f)c_2(f)g_1(f)}{G_{\tau_{(1),1}}(f;\theta)}-\sqrt{\frac{\alpha_1}{f}}g_1(f)-\gamma_1{\frac{\alpha_1}{f}}g'_1(f)\\
&+\sqrt{\frac{\alpha_1}{(M-1)f}}g_1((M-1)f)\frac{c_2((M-1)f)g_1((M-1)f)}
{1-G_{\tau_{(1),1}}((M-1)f;\theta)}\\
&+\sqrt{\frac{\alpha_1}{(M-1)f}}g_1((M-1)f)+\gamma_1{\frac{\alpha_1}{(M-1)f}}g'_1((M-1)f)\\
\end{align*}
Also
\begin{align*}
E \left[\frac{-\partial^2}{\partial\beta
\partial b}l(\theta)\right]&=\frac{1}{273+S_1}\frac{C\gamma_1}{2\beta}+\frac{nf}{\beta}\left\{[\frac{\alpha_1}{273+S_1}-\frac{\alpha_2}{273+S_2}].E_{\kappa}\right.\\
&\left.+M\frac{\alpha_1}{273+S_1}[1-G_{\tau_{(1),1}}((M-1)f;\theta)]+M\frac{\alpha_2}{273+S_2}G_{\tau_{(1),1}}((M-1)f;\theta)\right\}.
\end{align*}
Finally, we obtain
\begin{align*}
E \left[\frac{-\partial^2}{\partial\beta^2}l(\theta)\right]&=\frac{D}{\beta^2}+\frac{nf}{\beta^2}\left\{[\alpha_1-\alpha_2].E_{\kappa}\right.+M\alpha_1[1-G_{\tau_{(1),1}}((M-1)f;\theta)]\\
&\left.+M\alpha_2G_{\tau_{(1),1}}((M-1)f;\theta)\right\},
\end{align*}
where
\begin{align*}
D&=\sum_{k=2}^{M-1}\frac{\left[\Delta[\gamma_1\sqrt{\frac{\alpha_1}{kf}}g_1(kf)]\right]^2}
{\Delta[G_{\tau_{(1),1}}(kf;\theta)]}+\frac{\left[\gamma_1\sqrt{\frac{\alpha_1}{f}}g_1(f)\right]^2}{G_{\tau_{(1),1}}(f;\theta)}+
\frac{\left[\gamma_1\sqrt{\frac{\alpha_1}{(M-1)f}}g_1((M-1)f)\right]^2}{1-G_{\tau_{(1),1}}((M-1)f;\theta)}\\
&+\sum_{k=2}^{M-1}\left[\Delta[2\gamma_1\sqrt{\frac{\alpha_1}{kf}}g_1(kf)+g'_1(kf)(\gamma_1\sqrt{\frac{\alpha_1}{kf}})^2]\right]
+2\gamma_1\sqrt{\frac{\alpha_1}{f}}g_1(f)+g'_1(f)(\gamma_1\sqrt{\frac{\alpha_1}{f}})^2\\
&-2\gamma_1\sqrt{\frac{\alpha_1}{(M-1)f}}g_1((M-1)f)-g'_1((M-1)f)(\gamma_1\sqrt{\frac{\alpha_1}{(M-1)f}})^2\\
\end{align*}


\section*{Acknowledgment}

The authors would like to thank Professor M. N. Tata for her comments and suggestions which improved the first version of this paper.


\begin{thebibliography}{1}

\bibitem{be} M. Boulanger and L. A. Escobar. Experimental design for a
class of accelerated degradation tests. \emph{Technometrics}, 1994, {\bf 36}, 260 -� 272.\vspace{-0.2cm}

\bibitem{cz} Z. Chen and S. Zheng. Lifetime distribution based
degradation analysis. \emph{IEEE Trans. Reliability}, 2005, {\bf 54}, 3 -� 10.\vspace{-0.2cm}

\bibitem{hd} W. Huang and D. L. Dietrich. An alternative degradation
reliability modeling approach using maximum likelihood estimation.
\emph{IEEE Trans. Reliability}, 2005, {\bf 54},310 -� 317.\vspace{-0.2cm}

\bibitem{jy} V. R. Joseph and I. T. Yu. Reliability improvement
experiments with degradation data. \emph{IEEE Trans. Reliability}, 2006, {\bf 55}, 149 -� 157.\vspace{-0.2cm}

\bibitem {lc} J. Lawless and M. Crowder. Covariates and random effects in a gamma process model with
application to degradation and failure. \emph{Lifetime Data Analysis}, 2004, {\bf 10}, 213 -� 227.\vspace{-0.2cm}

\bibitem{lt}  C. M. Liao and S. T. Tseng. Optimal design for step-stress accelerated degradation
tests. \emph{IEEE Trans. Reliability}, 2006, {\bf 55}, 59 -� 66.\vspace{-0.2cm}

\bibitem {me} W. Q. Meeker and L. A. Escobar. \emph{Statistical Methods for
Reliability Data}. New York: John Wiley \& Sons, 1998.\vspace{-0.2cm}

\bibitem {mel} W. Q. Meeker, L. A. Escobar, and C. J. Lu. Accelerated
degradation tests: Modeling and analysis. \emph{Technometrics}, 1998, {\bf 40},
89 -� 99.\vspace{-0.2cm}

\bibitem{pb} Z. Pan and N. Balakrishnan. Multiple-steps step-stress accelerated
degradation modeling based on Wiener and Gamma processes. \emph{Commun. Stat. - Simul. Comput.}, 2010,
{\bf 39}, 1384 -- 1402.\vspace{-0.2cm}

\bibitem{pp1} C. Park and W. J. Padgett. Accelerated degradation
models for failure based on geometric Brownian motion and gamma
process. \emph{Lifetime Data Analysis}, 2005, {\bf 11}, 511 -� 527.\vspace{-0.2cm}

\bibitem{tyx} L.C. Tang, G.Y. Yang, M. Xie. Planning of step-stress accelerated degradation test. \emph{RAMS}, 2004, 287 -� 292.\vspace{-0.2cm}

\bibitem {tw} S. T. Tseng and Z. C. Wen. Step-stress accelerated degradation
analysis of highly-reliable products. \emph{Journal of Quality
Technology}, 2000, {\bf 32}, 209 -� 216.\vspace{-0.2cm}

\bibitem{yt2} H. F. Yu and S. T. Tseng. Designing a
degradation experiment. \emph{Naval Research Logistics}, 1999, {\bf 46}, 689 -� 706.\vspace{-0.2cm}

\bibitem{tbt} S. T. Tseng, N. Balakrishnan, C. C. Tsai. Optimal
Step-Stress Accelerated Degradation Test Plan for Gamma Degradation
Process. \emph{IEEE Trans. Reliability}, 2009, {\bf 58}, 611 -- 618.\vspace{-0.2cm}
\end{thebibliography}
\end{document}